\documentclass[aps,showpacs,twocolumn,twoside]{revtex4}
\usepackage{epsfig,amsmath,amssymb,graphicx,bm,psfrag}
\usepackage{dcolumn}
\usepackage{psfrag}

\newcommand{\ba}{\begin{eqnarray}}
\newcommand{\ea}{\end{eqnarray}}
\newcommand{\be}{\begin{equation}}
\newcommand{\ee}{\end{equation}}
\newcommand{\bmath}{\begin{mathletters}}
\newcommand{\emath}{\end{mathletters}}
\newcommand{\ban}{\begin{eqnarray*}}
\newcommand{\ean}{\end{eqnarray*}}

\newcommand{\bsub}{\begin{subequations}}
\newcommand{\esub}{\end{subequations}}

\newcommand{\reffig}[1]{\mbox{Fig.~\ref{#1}}}

\newcommand{\refsec}[1]{\mbox{Sec.~\ref{#1}}}

\begin{document}

\title{Long-range correlations in rectangular cavities containing point-like perturbations}

\author{Ma{\l}gorzata Bia{\l}ous}
\affiliation{Institute of Physics, Polish Academy of Sciences, Al. Lotnik\'ow 32/46, 02-668 Warszawa, Poland}
\author{Vitalii Yunko}
\affiliation{Institute of Physics, Polish Academy of Sciences, Al. Lotnik\'ow 32/46, 02-668 Warszawa, Poland}
\author{Szymon Bauch}
\affiliation{Institute of Physics, Polish Academy of Sciences, Al. Lotnik\'ow 32/46, 02-668 Warszawa, Poland}
\author{Micha{\l} {\L}awniczak}
\affiliation{Institute of Physics, Polish Academy of Sciences, Al. Lotnik\'ow 32/46, 02-668 Warszawa, Poland}
\author{Barbara Dietz}
\email{dietz@ifpan.edu.pl}
\affiliation{Institute of Physics, Polish Academy of Sciences, Al. Lotnik\'ow 32/46, 02-668 Warszawa, Poland}
\author{Leszek Sirko}
\email{sirko@ifpan.edu.pl}
\affiliation{Institute of Physics, Polish Academy of Sciences, Al. Lotnik\'ow 32/46, 02-668 Warszawa, Poland}

\begin{abstract}
We investigated experimentally the short- and long-range correlations in the fluctuations of the resonance frequencies of flat, rectangular microwave cavities that contained antennas acting as point-like perturbations. We demonstrate that their spectral properties exhibit the features typical for singular statistics. Hitherto, only the nearest-neighbor spacing distribution had been studied. We, in addition considered statistical measures for the long-range correlations and analyzed power spectra. Thereby, we could corroborate that the spectral properties change to semi-Poisson statistic with increasing microwave frequency. Furthermore, the experimental results are shown to be well described by a model applicable to billiards containing a zero-range perturbation [T. Tudorovskiy {\it et al.}, New. J. Phys. {\bf 12}, 12302 (2010)].
\end{abstract}


\date{\today}

\maketitle

\section{\label{intr}Introduction}
The field of quantum chaos~\cite{Chaos1991} focuses on the quantum manifestations of classical chaos. Signatures of chaoticity were observed in the statistical properties of the eigenvalues and the wave functions of the corresponding quantum system~\cite{Stoeckmann2000,Haake2001}, in the fluctuation properties of the scattering matrix elements of chaotic scattering processes~\cite{Verbaarschot1985,Guhr1998}, in the transport properties of quantum dots~\cite{Beenakker1997} and also in systems, where time-reversal invariance is broken, e.g., by a magnetic field~\cite{Berry1986}.  For a generic quantum system with classically regular dynamics the spectral properties were predicted to coincide with those of Poissonian random numbers~\cite{Berry1977}, whereas according to the Bohigas-Giannoni-Schmit conjecture~\cite{Bohigas1984} the spectral properties of chaotic systems are well described by random matrix theory (RMT)~\cite{McDonald1979,Casati1980,Berry1981,Mehta1990}. These predictions have been confirmed in a huge amount of experimental, numerical and theoretical studies by now. In the present article we focus on long-range correlations of the spectral fluctuations in non-chaotic systems. For this we used a procedure which is based on methods from time series analysis~\cite{Relano2002}. Within this approach, the $\delta_q$ statistic, defined as the deviation between the spacing of two unfolded levels separated by $(q-1)$ levels and their mean $q$, is viewed as a time series with the index $q$ taking the role of time. Its power spectrum $P_k^q$, obtained as the modulus square of its Fourier transform from $q$ to $k$ space, exhibits for small $k$ a power law $1/k^\alpha$, which allows to distinguish between a chaotic ($\alpha =1$) and an integrable ($\alpha =2$) classical dynamics. The full functional forms of the power spectrum were derived for the RMT ensembles and Poisson statistics in Ref.\cite{Faleiro2004}. The analytical result for time-reversal invariant chaotic systems, the spectral properties of which are expected to coincide with those of the eigenvalues of random matrices from the Gaussian orthogonal ensemble (GOE), were verified using a microwave cavity with the shape of a Sinai billiard~\cite{Faleiro2006}.

Generally, billiards provide appropriate systems for the study of problems within the field of quantum chaos, because the degree of chaoticity of their classical dynamics only depends on their shape~\cite{Chaos1991,Stoeckmann2000,Haake2001}. Furthermore, quantum billiards have the particular property that their eigenvalues and wave functions can be obtained experimentally by using flat, cylindrical microwave resonators~\cite{Stoeckmann1990,Sridhar1991,Graef1992,Sirko1997,Blumel2001,Hlushchuk2001}. Indeed, below a certain frequency $f_{\rm max}=c/(2h)$ with $c$ the velocity of light and $h$ the height of the resonator, the electrical field is perpendicular to the top and the bottom plate of the resonator and is governed by a two-dimensional Helmholtz equation with Dirichlet boundary conditions at the side walls of the resonator. Accordingly, in this frequency range of transversal-magnetic TM$_0$ modes it is mathematically equivalent to the Schr\"odinger equation of a quantum billiard of corresponding shape. Therefore, such resonators are referred to as microwave billiards. It is worth pointing out that microwave networks which simulate quantum graphs~\cite{Kottos1997,Kottos1999,Pakonski2001,Hul2004,Lawniczak2008,Lawniczak2010,Hul2012,Lawniczak2014} provide another suitable system for theoretical and experimental studies of problems within the field of quantum chaos.

We report on the investigation of long-range correlations of the spectral fluctuations in terms of power spectra in microwave billiards simulating singular billiards~\cite{Seba1990}, that is, quantum billiards which contain a point-like (zero-range) perturbations. In fact, microwave power is coupled into and out of the resonator via wire antennas~\cite{Haake1991}, and these act as singular scatterers. The cavities used in our experiments where rectangular. Such systems offer the simplest realization of a singular billiard, because the eigenvalues and the wave functions of the corresponding quantum billiard are known explicitly. There, indeed exist several experimental and theoretical studies on their spectral properties~\cite{Haake1991,Shighehara1993,Shighehara1994,Shighehara1996,Cheon1996,Weaver1995,Legrand1997,Rahav2002}. These works restrict to the investigation of the distribution of the spacings of adjacent energy levels, i.e., of short-range correlations. Already the very first experiments~\cite{Stoeckmann1990} revealed deviations of the nearest-neighbor spacing distribution (NNSD) from the Poissonian one, expected for integrable systems, which instead was of intermediate type~\cite{Bogomolny1998,Bogomolny2001}. For small spacings the NNSD vanished, i.e., exhibited level-repulsion typical for chaotic systems, whereas it decreased exponentially like the Poissonian level-spacing distribution for large spacings. These features of the NNSD were attributed to the presence of the antennas which act as singular perturbations. Yet, at first, these findings were surprising, because the classical dynamics of singular billiards is not chaotic. In classical mechanics point-like perturbations affect only those trajectories that hit them, and these are of measure zero. In contrast to the features, in quantum mechanics even small-size perturbations change the spectral properties of such non-chaotic systems. Note, however, that the spectral properties of billiards with a chaotic classical dynamics are not affected by the addition of singular perturbations.

In Refs.~\cite{Bogomolny2001a,Bogomolny2002} the two-point correlation function and the NNSD of integrable billiards with a $\delta$-function potential exhibiting a singular spectral statistics were derived. In Ref.~\cite{Exner1997} the microwave billiard with the attached antennas was regarded as a scattering system. This idea was readopted in Refs.~\cite{Tudorovskiy2008,Tudorovskiy2010,Tudorovskiy2011} and a rigorous equation for the computation of the energy levels was derived, applicable to microwave billiards with the shape of a classically integrable billiard.

The objective of the present article is the investigation of long-range correlations, such as the Dyson-Mehta statistic and power spectra of systems exhibiting singular statistics. In~\refsec{Experiment} the experiment is described and the experimental results are presented. Based on the Porter-Rosenzweig model~\cite{Rosenzweig1960} we determine the chaoticity parameter in~\refsec{Chaoticity} and, finally, in~\refsec{Singular} we compare the experimental results for the spectral properties to those of the eigenvalues of singular billiards. For this purpose, we computed the latter using the method described in Ref.~\cite{Tudorovskiy2010}. We will demonstrate, that with increasing microwave frequency the spectral properties of the microwave billiards approach semi-Poisson statistics~\cite{Tudorovskiy2010}.

\section{Experimental results\label{Experiment}}
\subsection{Experimental setup}

\begin{figure}[h!]
\centering
{\includegraphics[width=\linewidth]{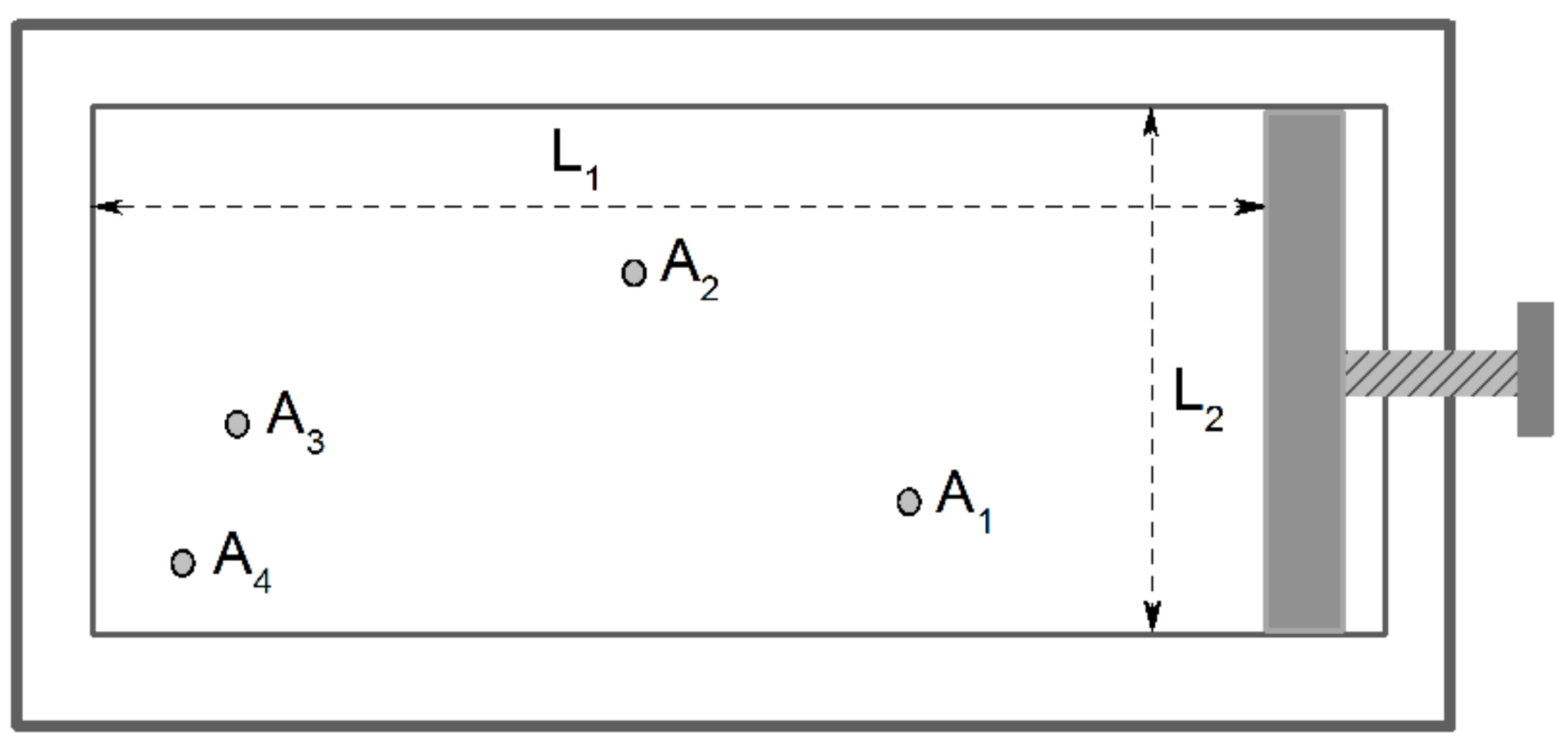}}
\caption{Schematic view of the rectangular microwave billiard. The length of its shorter side was kept fixed at L$_2=20.2$~cm, while the longer one took 5 different values differing by $0.2$~cm between L$_1=45.9$~cm and L$_1=46.7$~cm. For the transmission and reflection measurements, two microwave antennas were introduced into the cavity at two of the positions marked by A$_1$, A$_2$, A$_3$ and A$_4$.}
\label{fig1}
\end{figure}
\begin{figure}[h!]
\centering
{\includegraphics[width=\linewidth]{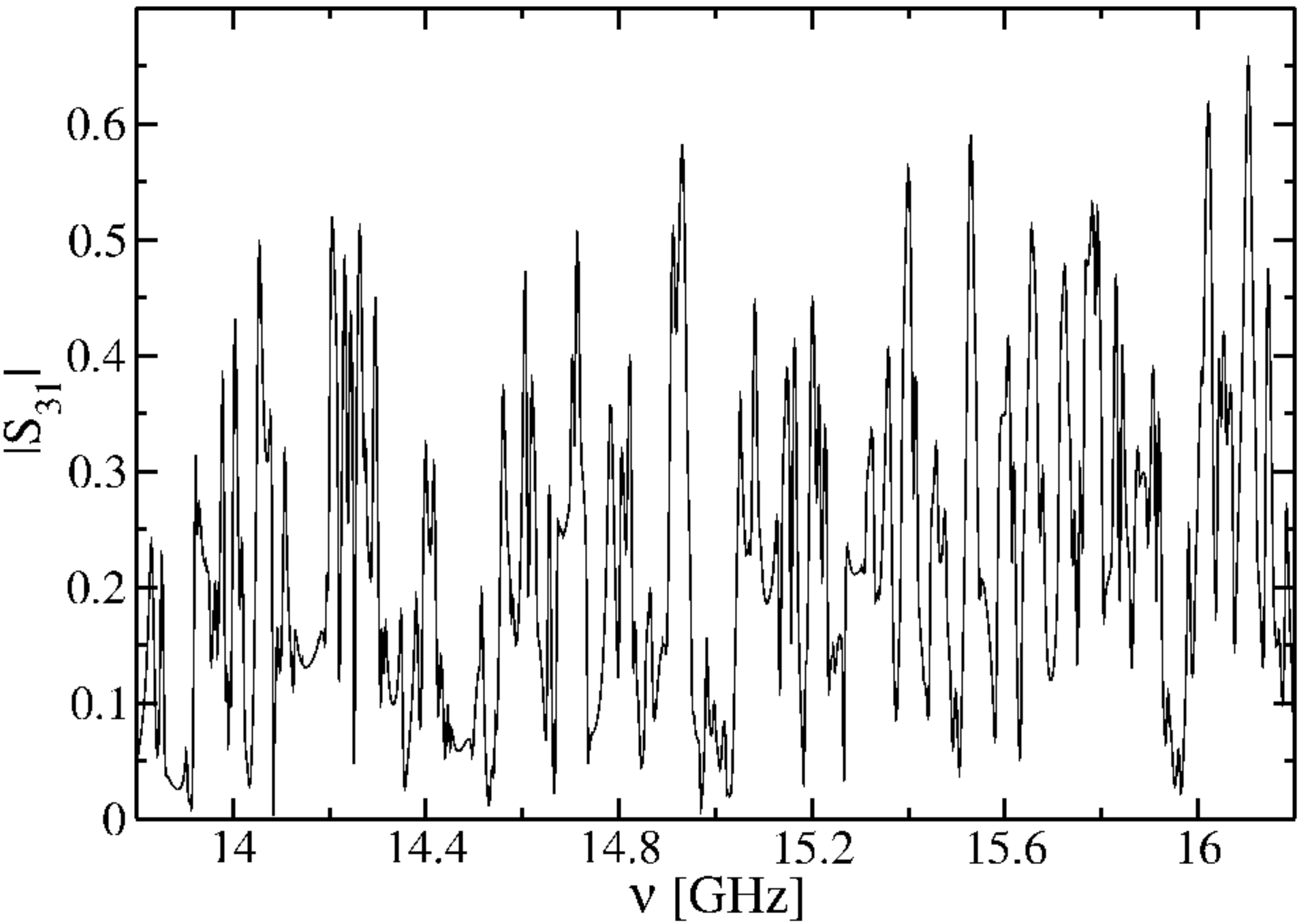}}
\caption{Transmission measurement between antennas positioned at A$_1$ and A$_3$ (see Fig.~\ref{fig1}) for the frequency range $13.5$~GHz$~\leq\nu\leq 16.5$~GHz. The spectrum consists of overlapping resonances. Therefore, in order to determine the resonance frequencies, we had to compare all reflection and transmission measurements for each resonator geometry. }
\label{fig2}
\end{figure}

The experiments were performed with a rectangular microwave billiard, a schematic view of which is shown in~\reffig{fig1}. The cavity was manufactured from brass and its height was $h=8$~mm corresponding to $f_{max}=18.6$~GHz. The measurements were performed up to $17$~GHz, in order to ensure the equivalence of the associated Helmholtz equation and the Schr\"odinger equation of the corresponding quantum billiard. The shorter side length L$_2$ of the microwave billiard was kept fixed at $20.2$~cm, whereas the longer one, L$_1$, was changed as indicated in~\reffig{fig1}, taking 5 values between $45.9$~cm to $46.7$~cm differing by $0.2$~cm. Two microwave antennas were introduced into the cavity at two of the positions marked by A$_1$, A$_2$, A$_3$ and A$_4$ in~\reffig{fig1}, for the measurements. With respect to the lower left corner they are located at $(31.6,5.2)$~cm, $(21.5,14.1)$~cm, $(5.7,8.2)$~cm and $(3.6,2.9)$~cm, respectively. The diameter of the wire was $0.9$~mm and it penetrated $3$~mm into the cavity. The resonance frequencies were obtained from transmission and reflection measurements. For this, microwave power was coupled into the resonator via one of the antennas and coupled out via the second or the same one by a vector network analyzer (Agilent E8364b) which was connected to the antennas through flexible microwave cables (HP 85133-616). A part of the transmission measurement with the antennas attached to the cavity at A$_1$ and A$_3$ is presented in~\reffig{fig2} in the frequency range from $13.5$~GHz to $16.5$~GHz.
It exhibits overlapping resonances, which hampered the determination of the resonance frequencies. The broadening of the resonances is mainly due to Ohmic absorption in the walls of the cavity. Actually, the quality factor $Q$ ranged between $Q\simeq 1000-4000$. In order to decide whether a hump in a broad resonance corresponds to a genuine eigenvalue we (i) plotted for each antenna combination the level sequences determined for the 5 configurations versus the discrete parameter values given by the lengths $L_1$ (see Fig.~\ref{fig1}), yielding a gapless level dynamics~\cite{Haake2001} if no levels are missing, (ii) compared the reflection and transmission measurements for a given configuration, (iii) plotted the fluctuating part of the integrated level density $N^{fluc}(\nu_i)$, that is, the difference of the number of identified resonance frequencies $N(\nu_i)=i$ below $\nu_i$ with $\nu_{1}\leq\nu_{2}\dots$ and the number predicted by Weyl's formula~\cite{Weyl1912} for microwave billiards, versus $\nu _i$. In particular this quantity provides a tool which is extremely sensitive to a missing or a spurious eigenvalue, because it exhibits jumps at such frequencies. Note that, if the electric field vanishes at the position of an antenna at a certain resonance frequency, then neither a resonance is excited nor a hump shows up at this frequency in the corresponding spectrum, resulting in a missing eigenvalue. While in the lowest frequency range, the six resulting sets of resonance frequencies are barely distinguishable, this is no longer the case in the upper frequency range, where the influence of the antennas becomes particularly perceptible, e.g., in the spectral properties. This aspect, actually, will be the subject of~\refsec{Singular}. Furthermore, the distance between the paremeters was sufficiently large, to ensure that the associated level sequences could be considered as independent, however, small enough so that we could identify missing levels as described above. Accordingly, in the lowest frequency range we had five independent level sequences, whereas in the upper freqency ranges the six combinations of antenna positions and the five parameters yielded 30 sets of independent level sequences.   

\subsection{Spectral properties of the microwave billiards}

Prior to the analysis of the spectral statistics, system specific properties, that is, the smooth modulations of the level density, need to be removed. We carried out this 'unfolding' by replacing the resonance frequencies $\nu_i$ by the smooth part of the integrated level density, which is given by Weyl's formula,
\begin{equation}
\epsilon_i =N^{smooth}(\nu_i).
\end{equation}
Weyl's formula~\cite{Weyl1912} corresponds to a quadratic polynomial, which depends on the area, the perimeter and the curvature of the billiard. On the one hand we used Weyl's formula for the unfolding, on the other hand, $N^{smooth}(\nu_i)$ was determined by fitting a quadratic polynomial to the experimentally determined integrated resonance density. Both procedures yielded the same results for the spectral properties. This yields dimensionless eigenvalues $\epsilon_i$ with mean value unity, $\langle s\rangle =1$, of the spacings $s_i=\epsilon_{i+1}-\epsilon_i$ between adjacent levels. 

We investigated the spectral properties of the unfolded resonance frequencies in terms of the NNSD $P(s)$, the Dyson-Mehta statistic $\Delta_3(L)$ and the power spectrum $\langle P_k^q\rangle$.  While the NNSD gives information on short-range correlations, the $\Delta_3$ statistic corresponds to a measure for the long-range correlations. The  $\Delta_3$ statistic is defined as the least-squares deviation of the integrated resonance density of the unfolded eigenvalues from the straight line best fitting it in the interval $L$ and provides a measure for the degree of rigidity of a level sequence. Another measure for long-range correlations is the $\delta_q$ statistic,
\begin{equation}
\label{deltaq}
\delta_q=\sum_{i=1}^{q}(s_i-\langle s \rangle) = \epsilon_{q+1}-\epsilon_1-q,
\end{equation}
which gives the deviation of the spacing between two unfolded levels separated by $(q-1)$ levels from its mean $q$. Considering the index $q$ as the analogue of a discrete time, the power spectrum is obtained as the modulus square of the Fourier transform from 'time' to $k$ space~\cite{Relano2002},
\begin{equation}
\label{Sq}
P_k^q=|\tilde{\delta}_k |^2,
\end{equation}
where $\tilde{\delta}_k$ is the Fourier transform of $\delta_q$,
\begin{equation}
\label{FFTdeltaq}
\tilde{\delta}_k=\frac{1}{\sqrt{N}}\sum_{q=0}^{N-1}\delta_q\exp\left(-2\pi iq\frac{k}{N}\right).
\end{equation}

The analytical results for the NNSD and the Dyson-Mehta statistic are given by~\cite{Mehta1990}
\begin{eqnarray}
P^{\rm Poisson}(s)&=&e^{-s},\\
P^{\rm GOE}(s)&\simeq&\frac{\pi}{2}se^{-\frac{\pi}{4}s^2},
\label{NND}
\end{eqnarray}
and
\begin{eqnarray}
\Delta_3^{\rm Poisson}(L)&=&\frac{L}{15},\\
\Delta_3^{\rm GOE}(L)&\simeq&\frac{1}{\pi^2}\left(\ln(2\pi L)+\gamma
-\frac{5}{4}-\frac{\pi^2}{8}\right),
\label{Delta3}
\end{eqnarray}
respectively.  Here, $\gamma=0.5772...$ is Euler's constant. The quoted GOE results are approximations which, however, have been shown to describe the spectral properties of the eigenvalues of random matrices from the GOE very well~\cite{Mehta1990}. In Ref.~\cite{Faleiro2004} analytical expressions where derived on the basis of RMT for the Gaussian ensembles and for Poissonian random numbers, yielding
\begin{eqnarray}
\label{Analytical}
\langle P_k^q\rangle &=&\nonumber
\frac{1}{4\pi^2}\left[\frac{K(k/N)-1}{\left(k/N\right)^2} + \frac{K(1-k/N)-1}{(1-k/N)^2}\right]\\
&+& \frac{1}{4\sin^2(\pi k/N)} + \Delta.
\end{eqnarray}
Here, $\Delta = -1/12$ for the Gaussian ensembles and $\Delta = 0$ for Poisson sequences and $K(\tau)$ is the spectral form factor,
\begin{eqnarray}
K^{Poisson}(\tau )&=&1\\
K^{GOE}(\tau)&=&2\tau -\tau\ln\left(1+2\tau\right).
\label{Ktau}
\end{eqnarray}
Note, that in Eq.~(\ref{Analytical}) the variable $\tau=k/N$ takes values between $0<1/N\leq\tau\leq 1-1/N <1$.
For $k/N\ll 1$ the average power spectrum approaches a power-law behavior $\langle P_k^q \rangle\propto 1/(k/N)^\alpha$ which can be summarized as follows,
\begin{equation}
\label{Asymptotic}
\langle P_k^q\rangle\longrightarrow
\begin{cases}
\frac{1}{4\pi^2(k/N)^2}, &\mbox{Poisson statistic},\\
\frac{1}{2\pi^2 (k/N)}, &\mbox{GOE}.\\
\end{cases}
\end{equation}
Thus, already the asymptotic features of  $\langle P_k^q \rangle$ for $k\rightarrow 0$ provide information on the chaoticity of the underlying classical dynamics.

\begin{figure}[h!]
\centering
{\includegraphics[width=\linewidth]{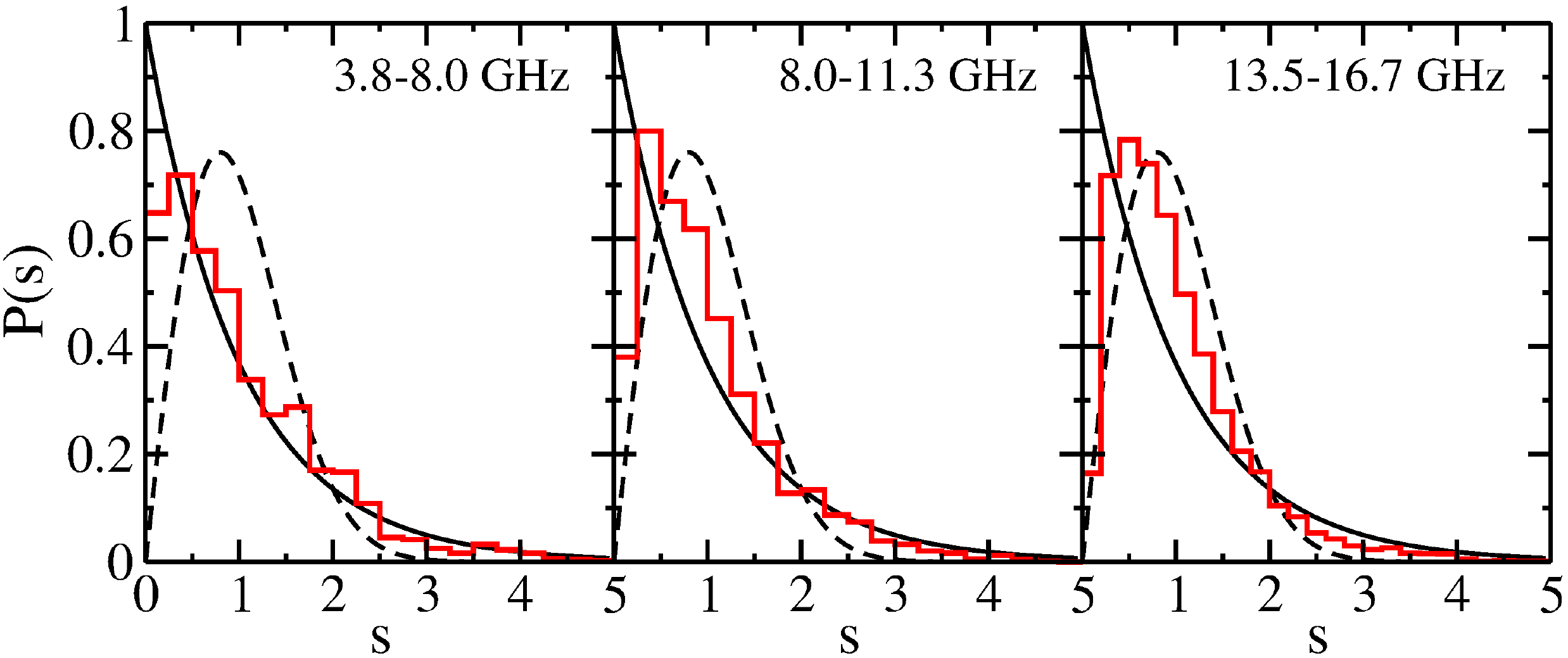}}
{\includegraphics[width=\linewidth]{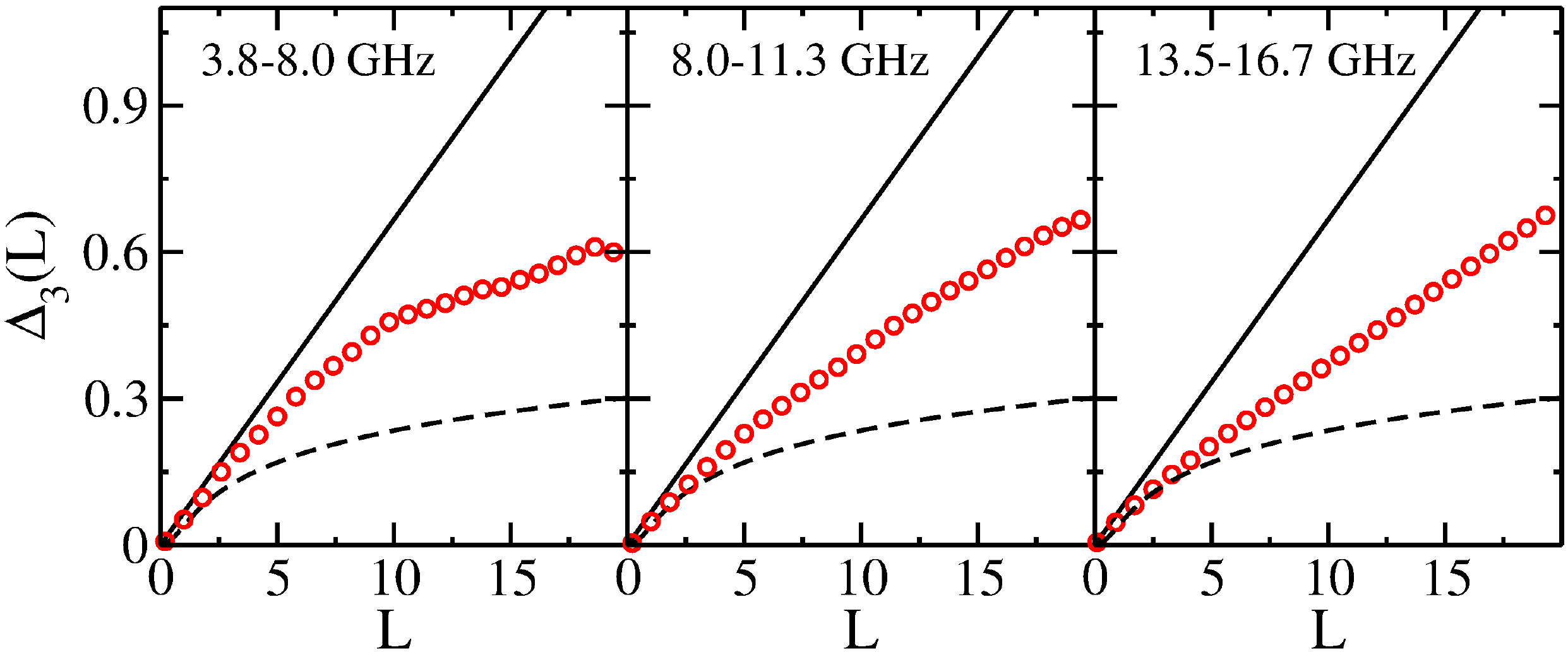}}
{\includegraphics[width=\linewidth]{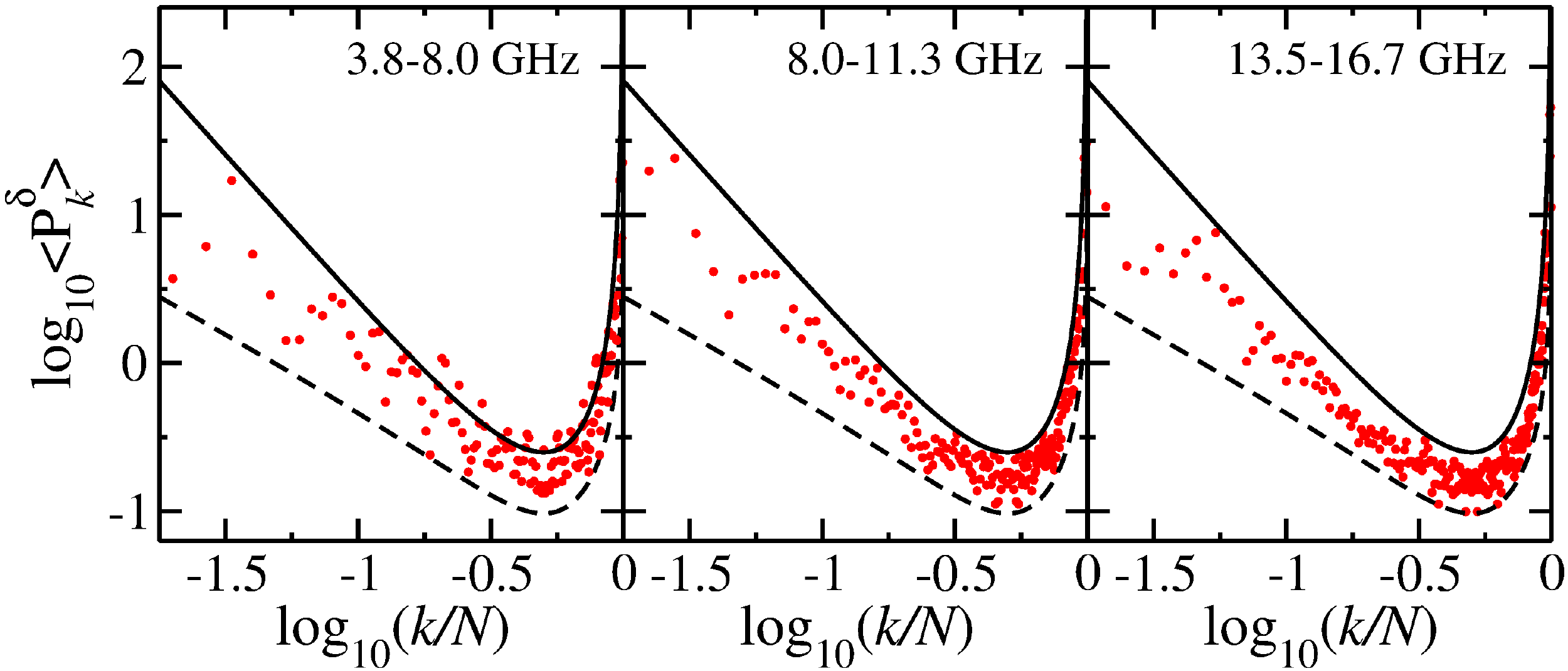}}
\caption{(Color online) Experimental results for the NNSD (red [gray] histograms in upper panel), the $\Delta_3$ statistic (red [gray] circles in the middle panels) and the power spectrum (red [gray] dots in lower panel). The corresponding curves for Poisson and GOE statistics are shown as full and dashed lines, respectively.}
\label{fig3}
\end{figure}

We realized that the spectral fluctuation properties of the experimental resonance frequencies varied smoothly with increasing frequency. Therefore, we investigated them for each of the 30 data sets in three frequency ranges, namely in the intervals [3.8,8.0]~GHz, [8.0,11.3]~GHz and [13.5,16.7]~GHz, corresponding to $150$, $180$ and $225$ resonance frequencies, respectively. The intervals were chosen such that the spectral properties were approximately the same over their whole range. The statistical measures were determined for each set separately and then averaged over the ensemble. For the calculation of the $\Delta_3$ statistic we proceeded as described in~\cite{Bohigas1983}. In ~\reffig{fig3} we present the thus obtained results for the NNSD (red [gray] histograms in the upper panels), the $\Delta_3$ statistic (red [gray] circles in the middle panels) and the power spectrum (red [gray] dots in the lower panels). They are compared to the corresponding Poisson and the GOE results, shown as full and dashed lines, respectively. In all frequency intervals the spectral properties clearly differ from Poisson, which is expected to describe the properties of billiards with integrable classical dynamics like rectangular billiards. We performed numerical calculations in order to ensure, that these deviations can not be attributed to the fact that the ratios of the side lengths L$_1$ and L$_2$ of the microwave billiards used in the experiments were no irrational numbers~\cite{Robnik1998}.

\begin{figure}[h!]
\centering
{\includegraphics[width=\linewidth]{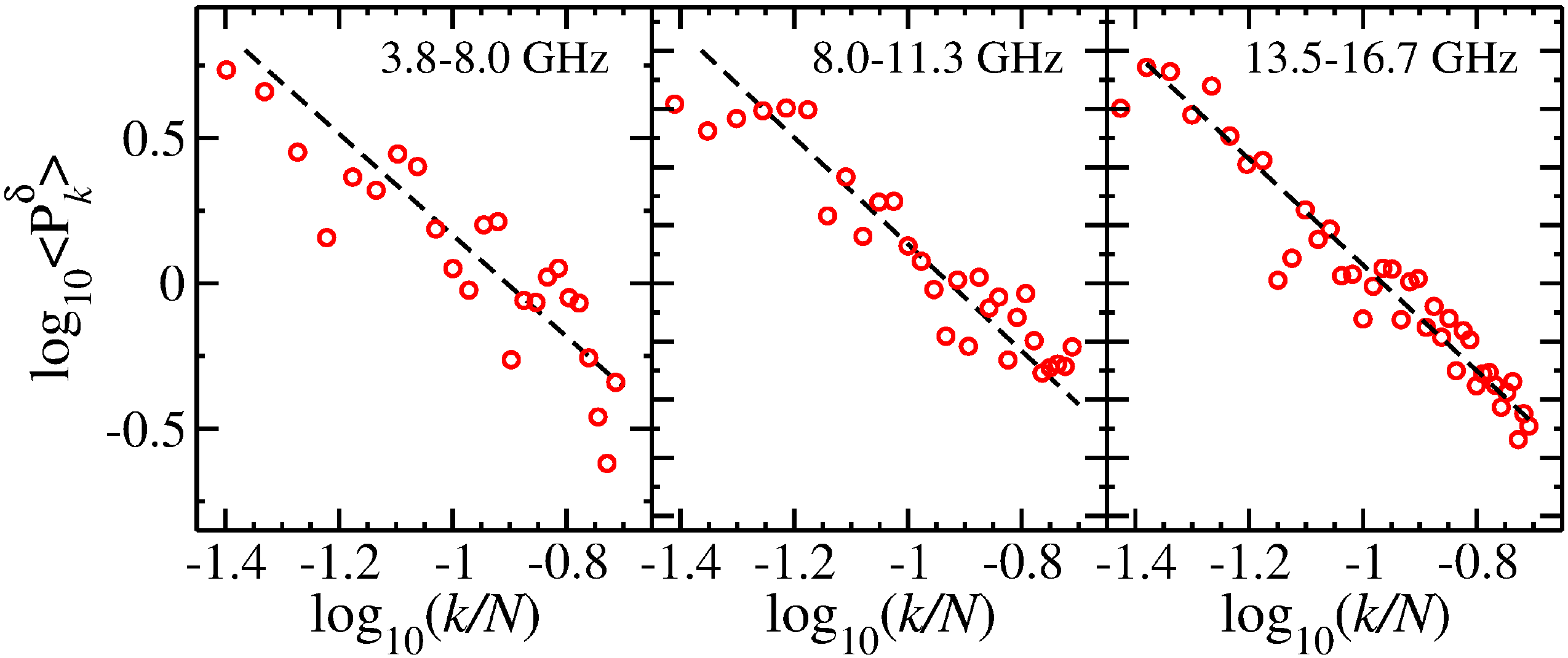}}
\caption{(Color online) Asymptotic behavior of the power spectra, shown in the lower panels of~\reffig{fig3}, in a log-log plot. The experimental results and the straight line best fitting the data are plotted as red (gray) circles and as dashed lines, respectively. The slopes of the latter are close to $-2$ in all frequency intervals.}
\label{fig3f}
\end{figure}

The agreement with Poisson becomes worse with increasing frequency. At $s=0$ the NNSD has a dip for all frequency ranges and is vanishingly small for the uppermost interval. Thus it exhibits there the same features as generic chaotic systems. For large spacings, on the other hand, the NNSDs decrease exponentially like the Poissonian one. Also the $\Delta_3$ statistic and the power spectra neither agree with that for Poisson statistic nor with that for the GOE in all frequency intervals and seem to approach the corresponding GOE curve with increasing frequency. We determined the power-law behavior $\langle P_k^q \rangle\propto 1/(k/N)^\alpha$ illustrated in a log-log plot in~\reffig{fig3f} and found that $\alpha\simeq 1.745$ in the lower frequency range, $\alpha\simeq 1.836$ in the middle one and $\alpha\simeq 1.850$ in the upper interval. These values are close to $\alpha\simeq 2.0$, expected for Poissonian random numbers; see Eq.~(\ref{Asymptotic}). Note, however, that the
 smallest value of $k/N$ achieved in the experiments was $1/N\simeq 0.005-0.007$. This prohibited the determination of $\alpha$ in the effectively asymptotic region. Still these results indicate, that the fluctuation properties of the resonance frequencies of the microwave billiard exhibit a statistic intermediate between Poisson and GOE, thereby confirming previous findings~\cite{Haake1991,Exner1997,Lawniczak2015}. The aim of the following section is to quantify the deviation of the spectral properties from regularity in terms of a chaoticity parameter.

\section{Determination of the chaoticity parameter \label{Chaoticity}}

\begin{figure}[h]
\centering
{\includegraphics[width=\linewidth]{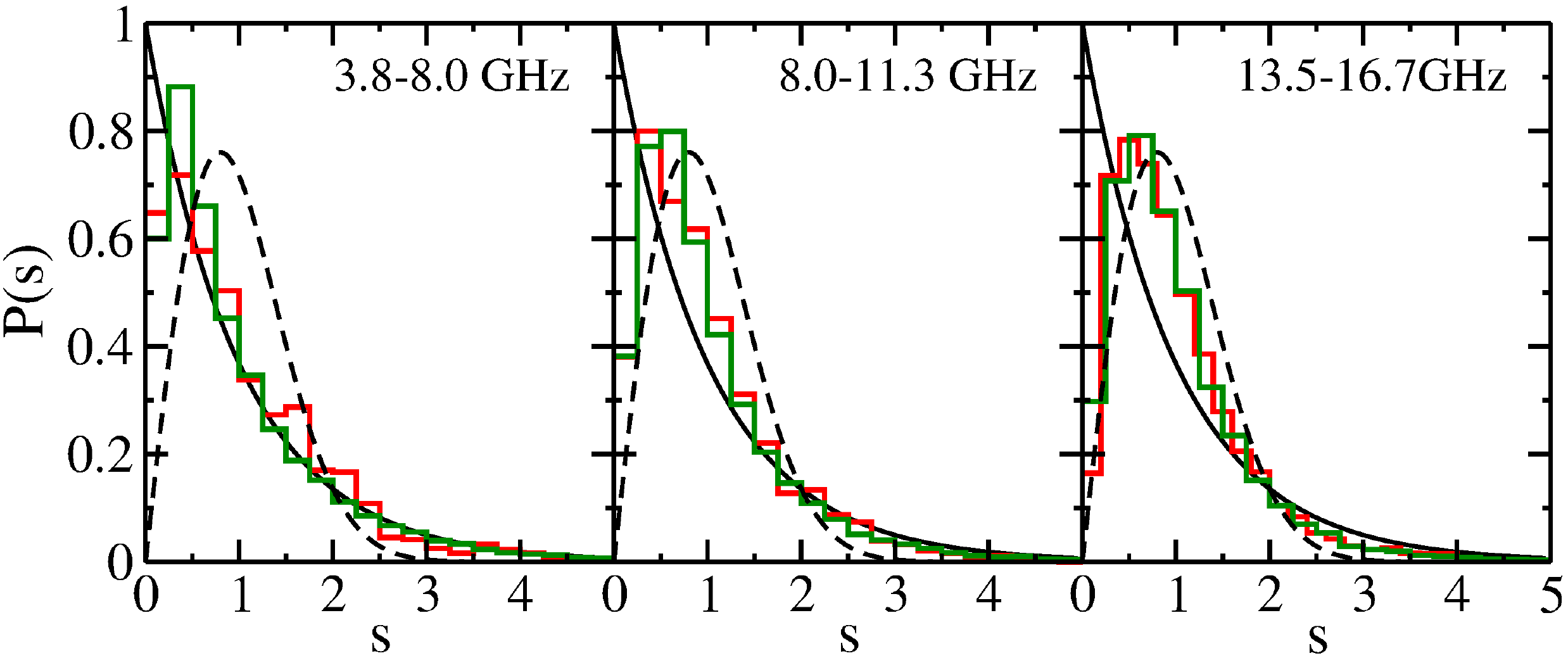}}
{\includegraphics[width=\linewidth]{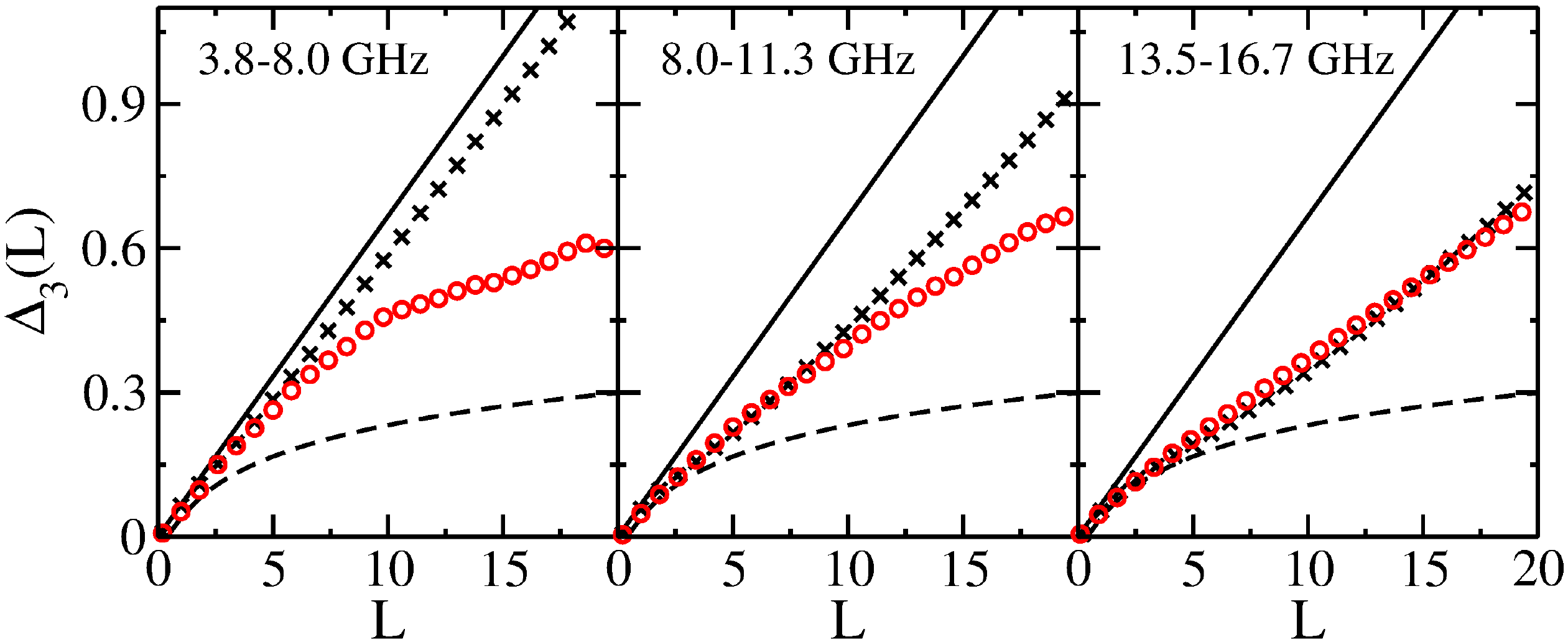}}
{\includegraphics[width=\linewidth]{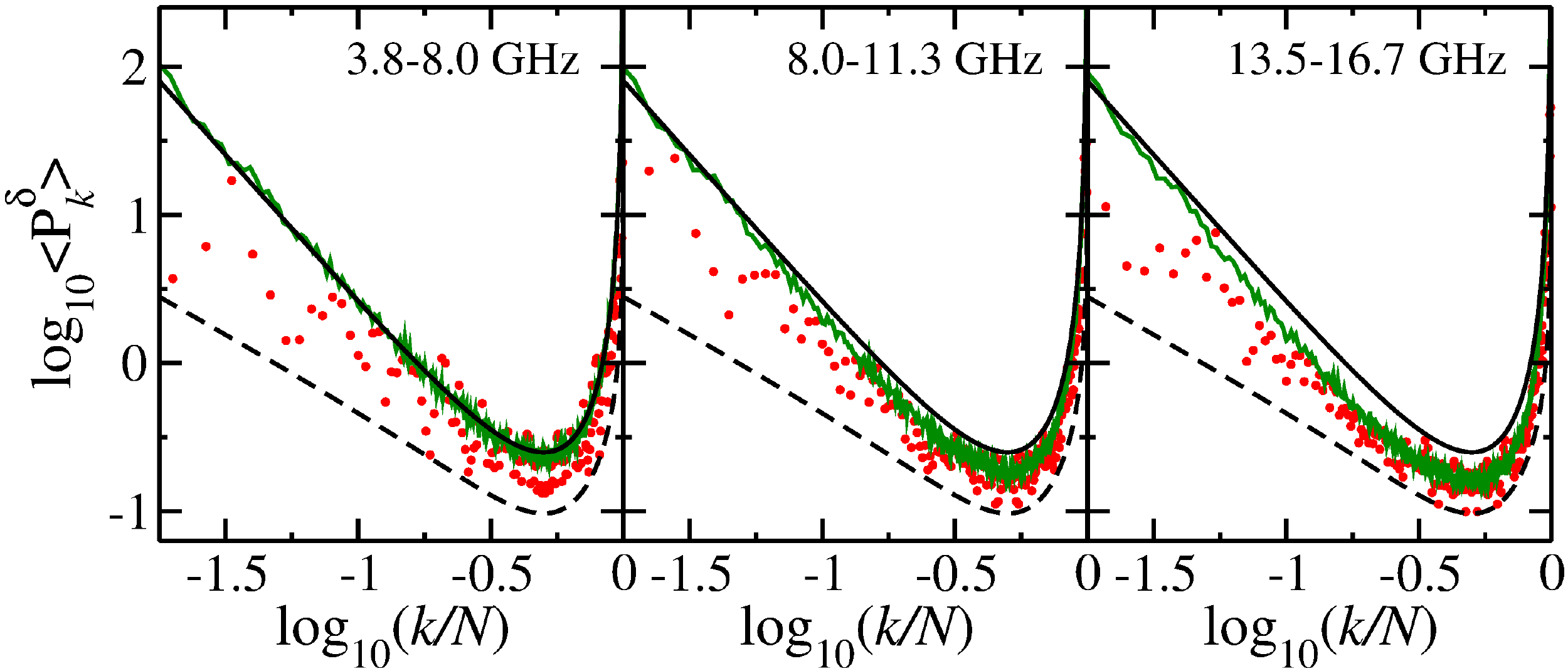}}
\caption{(Color online) Same as Fig.~\ref{fig3}. The experimental results are compared to the curves deduced from the Porter-Rosenzweig model, shown as green (light gray) lines in the upper and lower panels and as black crosses in the middle panel. The chaoticity parameter equals $\kappa =0.6$ in the frequency range 3.8 - 8.0~GHz, $\kappa =1.5$ for 8.0 - 11.3~GHz and $\kappa =2.25$ in the interval 13.5 - 16.7~GHz.}
\label{fig4}
\end{figure}

In order to estimate the size of the deviation of the spectral properties from those of Poissonian random numbers we applied the random-matrix model by Porter-Rosenzweig~\cite{Rosenzweig1960}, which depends on a parameter $\kappa$ and interpolates between Poisson for $\kappa =0$ and GOE for $\kappa =N$, with $N$ denoting the dimension of the random matrices $\hat H$ with matrix elements
\begin{equation}
\hat H_{ij}=\hat G_{ij}\left[\delta_{ij}+\frac{\kappa}{N}(1-\delta_{ij})\right],\, i,j=1,....,N.
\end{equation}
Here, the quantities $\hat G_{ij}$ denote the entries of a real symmetric matrix from the GOE. The parameter $\kappa$ is defined in such a way, that it does not depend on the dimension $N$ of $\hat H$. In order to verify this, we generated for various values of $\kappa$ ensembles of 500 random matrices with dimensions $N=200-500$, and computed the NNSD, the $\Delta_3$ statistic and the power spectra. Then we determined the values of $\kappa$ corresponding to the three frequency intervals by comparing the RMT and the experimental results for the statistical measures and by computing the corresponding mean square deviations, yielding $\kappa =0.6$ in the frequency range [3.8,8.0]~GHz, $\kappa =1.5$ for [8.0,11.3]~GHz and $\kappa =2.25$ for [13.5,16.7]~GHz. These values are in line with our observation that the statistical measures deviate more and more from Poisson with increasing frequency. In~\reffig{fig4} we show the experimental curves together with those obtained on the basis of the Porter-Rosenzweig model.
Especially in the uppermost frequency interval, the agreement between the experimental NNSD and the RMT result seems to be very good. However, the Porter-Rosenzweig model fails to reproduce the experimental long-range correlations, as illustrated for the $\Delta_3$ statistic and the power spectra shown in the middle and the lower panels of~\reffig{fig4}. Similar observations were made for the NNSD in Ref.~\cite{Haake1991}. Accordingly, we basically determined the values of $\kappa$ from the NNSD. Nevertheless, the parameter $\kappa$ provides a suitable measure for the size of the deviations of the spectral statistic from Poisson.

\section{Comparison with singular statistics \label{Singular}}

The deviations of the spectral properties of the microwave billiards from the expected Poissonian statistics are attributed to the presence of the antennas. Thin wire antennas can be described as two-dimensional dipoles (line sources) with a frequency dependent coupling to the resonator~\cite{Haake1991}. The corresponding Helmholtz equation is mathematically equivalent to the Schr{\"o}dinger equation of a singular quantum billiard~\cite{Bogomolny2001},
\begin{equation}
\label{singular_eigenvalues}
\left[-\Delta+V(\nu )\delta(\boldsymbol{x}-\boldsymbol{x}_0)\right]\Psi(\boldsymbol{x}) =k^2\Psi(\boldsymbol{x}),
\end{equation}
where $\boldsymbol{x}_0$ denotes the position of the antenna and the wave function $\Psi(\boldsymbol{x})$ has to fulfill the Dirichlet boundary condition. It was shown in~\cite{Haake1991} that the coupling parameter $V(\nu )$ varies slowly with frequency. This is in accordance with our findings, obtained by varying the lengths and the positions of the frequency intervals used for the analysis of the spectral properties, and allowed us to investigate the spectral properties in frequency intervals containing at least 150 resonance frequencies. In Refs.~\cite{Bogomolny2001a,Bogomolny2002} the two-point correlation function and the NNSD of the eigenvalues of this equation were derived for a fixed $V(\nu )$ and shown to exhibit a singular statistics intermediate between Poisson and GOE. This approach is applicable to closed singular billiards. Microwave billiards, however, are open systems with the antennas acting as scattering channels. In Ref.~\cite{Tudorovskiy2010} a model was developed which is applicable to such situations. The microwave billiard was regarded as a scattering system and the antennas as point-like scatterers and the computation of the eigenvalues was reduced to the calculation of the associated renormalized Green's function. This yielded an explicit eigenvalue equation for singular billiards with rectangular shape,
\begin{equation}
\sum_{n,m=1}^\infty\Psi^2_{n,m}(\boldsymbol{x}_0)\frac{e^{1-k^2_{nm}/k^2}}{k^2-k^2_{nm}}+\frac{1}{2\pi}\ln\left(\frac{k\beta}{2}\right)+\varkappa=0,
\label{eigenvalues}
\end{equation}
where $k^2_{nm}$ and $\Psi_{nm}$ are the eigenvalues and wave functions of the unperturbed rectangular billiard, and $\varkappa\simeq -0.058942$ is a constant. This equation depends only on one parameter, $\beta$, which can be interpreted as the scattering length of the perturbation~\cite{Tudorovskiy2010}. The perturbation is week for $\beta\geq 1$ and increases with decreasing $\beta <1$.

We determined 20000 eigenvalues of the unperturbed rectangular billiards for all geometries considered in the experiment and solved the eigenvalue problem Eq.~(\ref{eigenvalues}) for each antenna position $\boldsymbol{x}_0$ (see~\reffig{fig1}) for several values of $\beta$. Then we compared the spectral properties for the different realizations with the experimental ones and computed the corresponding mean square deviations, in order to determine the value of $\beta$ for which the model best describes the experimental results, yielding $\beta =0.9$ for the frequency range [0.1,8.0]~GHz, $\beta =0.5$ for [8.0,11.3]~GHz and $\beta =0.01$ for the interval [13.5,16.7]~GHz. For the analysis of the spectral properties we only took into account the perturbed eigenvalues, for which $\Psi_{nm}(\boldsymbol{x}_0)$ was nonzero. Note, that the unperturbed wave functions correspond to a vanishing electric field strength at the position of the antenna and, thus, the corresponding resonance frequencies are missing in the experimental eigenvalue list. Furthermore, we used a similar number of eigenvalues as was found in the respective frequency interval.

\begin{figure}[h]
\centering
{\includegraphics[width=\linewidth]{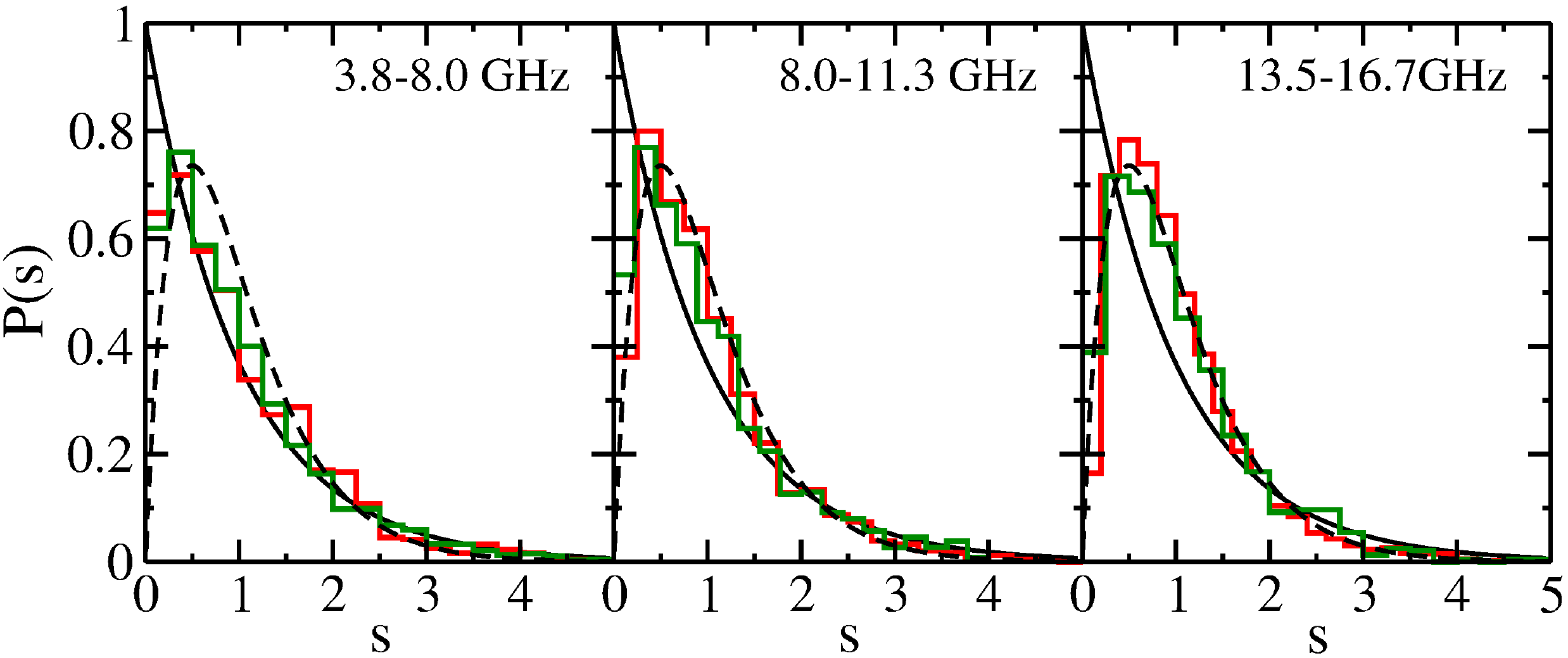}}
{\includegraphics[width=\linewidth]{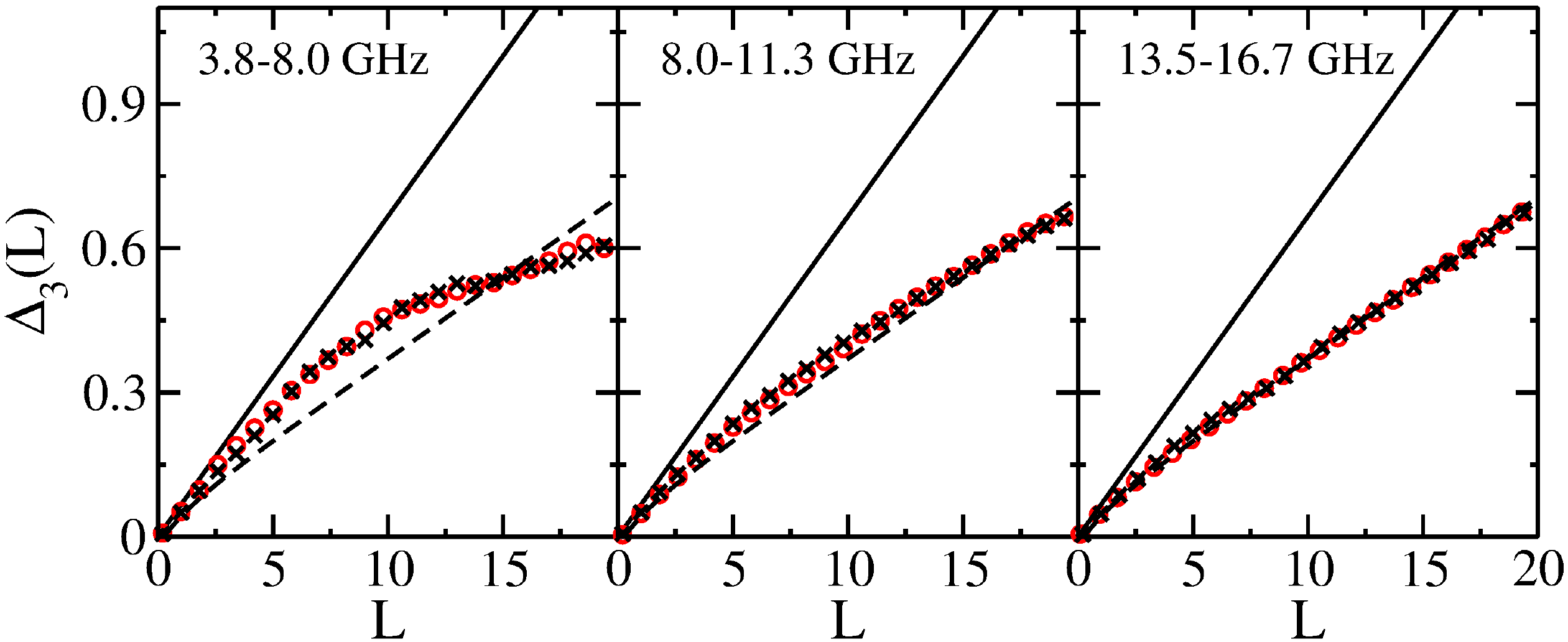}}
{\includegraphics[width=\linewidth]{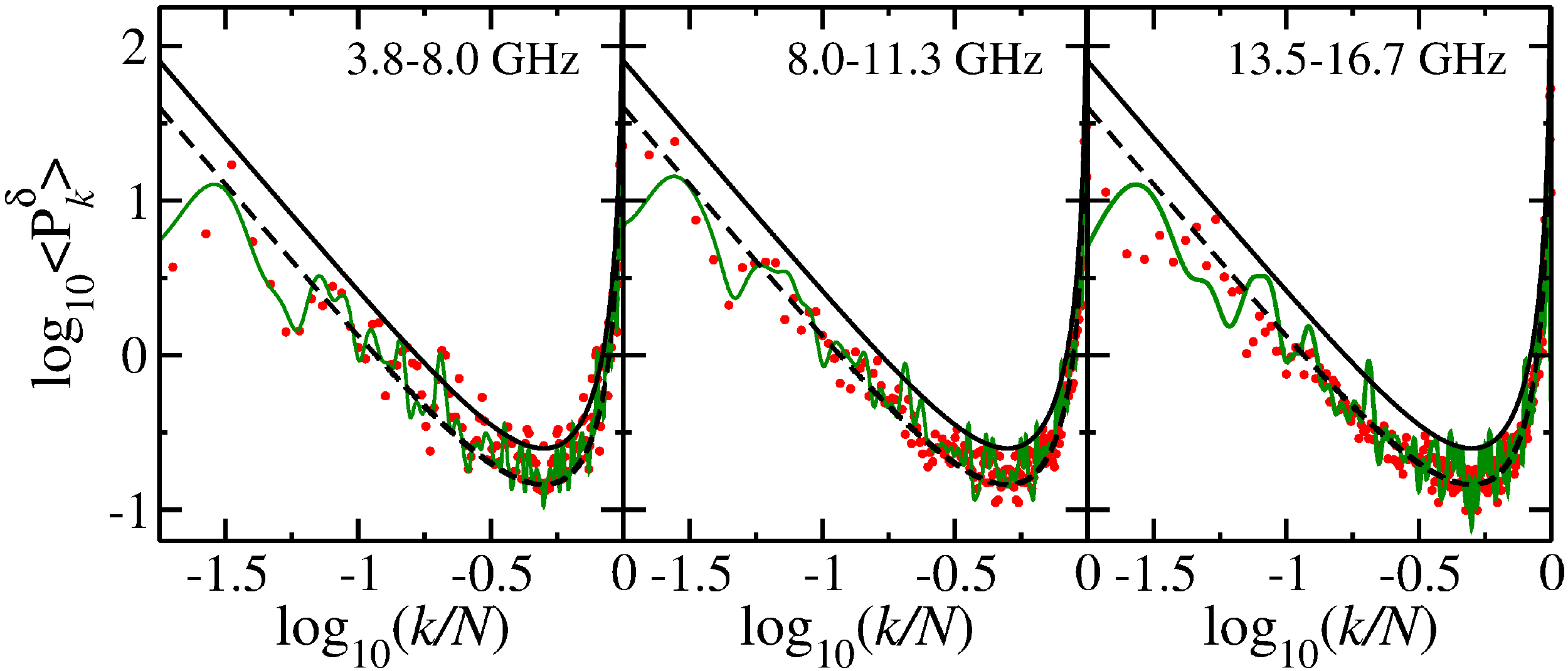}}
\caption{(Color online) Same as Fig.~\ref{fig3}. The experimental results are compared to the curves obtained for the corresponding singular billiards, shown as green (light gray) lines in the upper and lower panels and as black crosses in the middle panels, for the scattering length of the perturbation $\beta =0.9$ in the frequency range 3.8 - 8.0~GHz, $\beta =0.5$ for 8.0 - 11.3~GHz and $\beta =0.01$ in the interval 13.5 - 16.7~GHz. The corresponding results for Poisson and semi-Poisson statistics are shown as full and dashed lines, respectively. In the uppermost frequency interval the curves for semi-Poisson and the experimental and RMT results lie on top of each other.}
\label{fig5}
\end{figure}

In~\reffig{fig5} we compare the experimental results with the numerical ones, obtained by solving Eq.~(\ref{eigenvalues}) (green curves in the upper and the lower panels, crosses in the middle ones). The agreement is very good for all three statistical measures. Both results are compared to semi-Poisson statistics (dashed line), where the NNSD and the $\Delta_3$ statistic are given by~\cite{Bogomolny2001}
\begin{equation}
P^{sP}(s)=4se^{-2s}
\end{equation}
and
\begin{equation}
\Delta^{sP}_3(L)=\frac{L}{30}+\frac{1}{16}+\frac{e^{-4L}}{64l^2}\left[1+O\left(\frac{1}{L}\right)\right] +O\left(\frac{1}{L}\right),
\end{equation}
respectively. The result for the power spectrum is obtained by inserting the form factor~\cite{Bohigas2006}
\begin{equation}
K^{sP}(\tau)=\frac{2+\pi^2\tau^2}{4+\pi^2\tau^2}
\end{equation}
and $\Delta=-1/24$ into Eq.~(\ref{Analytical}). For small values of $k/N$ the power spectrum exhibits a power-law behavior  $\langle P_k^q \rangle\propto 1/(k/N)^\alpha$ with $\alpha\simeq 1.89$ in the experimental range of $k$ values (see~\reffig{fig3f}). This values is close to the experimental ones. The agreement of the experimental and the numerical results with semi-Poisson is striking in the uppermost frequency range. In the middle panels, showing the $\Delta_3$ statistic, all three curves actually lie on top of each other. A similar agreement with semi-Poisson was found numerically for the NNSD in Ref.~\cite{Tudorovskiy2010}.
In conclusion, the spectral properties of the rectangular microwave billiards are well described by the model for billiards containing zero-range perturbations.
Finally, we should note that the spectral statistics of the eigenvalues of Eq.~(\ref{eigenvalues}) changes towards Poisson for high energies $k^2_n$, i.e., far beyond the experimentally achievable values. Here, the value of $n\gg 1$ depends on the scattering length $\beta$, that is, the coupling parameter $V(\nu)$ in Eq.~(\ref{singular_eigenvalues}). 

\section{\label{Concl}Conclusions}

We investigated the short- and long-range correlations of the spectral fluctuation properties of rectangular microwave billiards, which contain antennas that act as point-like perturbations and showed that they exhibit singular statistics. Thus, such systems can be used to model the features of singular quantum billiards. The experimental data indeed are well described by a model applicable to singular rectangular billiards, i.e., to billiards containing zero-range perturbations~\cite{Tudorovskiy2010}. We observe a transition from Poisson statistics towards semi-Poisson with increasing frequency. Indeed, in the uppermost frequency range achieved in the experiments, the spectral properties are surprisingly well described by semi-Poisson statistics. In order to corroborate these observations, we did not only consider the NNSD, as had been done before, but also studied long-range correlations such as the $\Delta_3$ statistic and the $\delta_q$ statistic, or rather its Fourier
 transform, the power spectrum. In~\cite{Lawniczak2015} deviations of the elastic enhancement factor computed on the basis of transmission and reflection measurements in the frequency range [16,18.5]~GHz with rectangular microwave billiards, from the predictions for integrable systems were found which may be explained by these results. We, in fact, may conclude, that these spectra instead are suitable for the study of the enhancement factor of systems exhibiting singular statistics. This aspect will be pursued in a future work. Finally, we may conclude that in order to obtain a statistics close to Poisson, i.e., $\beta\simeq 1$ for the fluctuations in the spectra of microwave billiards with classically integrable dynamics, the lengths of the antennas need to be minimized. This will result in small excitations of the resonance modes, and thus small amplitudes in the corresponding transmission and reflection measurements. Then, however, the identification of the resonance positions, and thus the determination of the eigenvalues, will become even more cumbersome if not impossible, yielding incomplete spectra. Strictly speaking, due to the presence of the antennas the realization of Poisson statistics is impossible in microwave billiards. Nevertheless, a level statistics which is indistinguishable from Poissonian has been achieved for the first few hundreds of resonance frequencies in measurements with superconducting microwave billiards with an integrable classical dynamics. In these experiments, the lengths of the antennas was chosen such that they did not reach into the cavity~\cite{Dietz2014,Dietz2015,Dietz2015a}. Thereby, any effect of the antennas on the spectral properties was reduced considerably and at the same time all eigenmodes could be excited sufficiently to obtain complete level sequences.   

\begin{acknowledgments}
This work was partially supported by the Ministry of Science and Higher Education grant No. UMO-2013/09/D/ST2/03727 and the EAgLE project (FP7-REGPOT-2013-1, Project Number: 316014).
\end{acknowledgments}


\begin{thebibliography}{1}

\bibitem{Chaos1991} \emph{Chaos and Quantum Physics}, edited by M. J. Giannoni, A. Voros, and J. Zinn Justin (North- Holland, N.Y., 1991).
\bibitem{Stoeckmann2000} H.-J. St{\"o}ckmann, \emph{Quantum Chaos: An Introduction} (Cambridge University Press,
\bibitem{Haake2001} F. Haake, \emph{Quantum Signatures of Chaos} (Springer-Verlag, Heidelberg, 2001).
\bibitem{Verbaarschot1985} J.~Verbaarschot, H.~Weidenm\"uller, and M.~Zirnbauer, Phys. Rep. {\bf 129}, 367 (1985).
\bibitem{Guhr1998} T.~Guhr, A.~M\"{u}ller-Groeling, and H.~A. Weidenm\"{u}ller, Phys. Rep. {\bf 299}, 189 (1998).
\bibitem{Beenakker1997} C.~W. Beenakker, Rev. Mod. Phys. {\bf 69}, 731 (1997).
\bibitem{Berry1986} M.~V. Berry and M.~Robnik, J. Phys. A {\bf 19}, 649 (1986).
\bibitem{Berry1977} M.V. Berry and M. Tabor, Proc. R. Soc. A {\bf 356}, 375 (1977).
\bibitem{Bohigas1984} O. Bohigas,M.~J. Giannoni, and C. Schmit, Phys. Rev. Lett. {\bf 52} (1984).
\bibitem{McDonald1979} S.~W. McDonald and A.~N. Kaufman, Phys. Rev. Lett. {\bf 42}, 1189 (1979).
\bibitem{Casati1980} G. Casati, F. Valz-Gris, and I. Guarnieri, Lett. Nuovo Cimento {\bf 28} (1980).
\bibitem{Berry1981} M.~V. Berry, Eur. J. Phys. {\bf 2}, 91 (1981).
\bibitem{Mehta1990} M.~L. Mehta, \emph{Random Matrices} (Academic Press, London, 1990).
\bibitem{Relano2002} A. Rela\~no, J.M.G. Gomez, R. A. Molina, J. Retamosa, Phys. Rev. Lett. {\bf 89}, 244102 (2002).
\bibitem{Faleiro2004} E. Faleiro, J. M. G. G\'omez, R. A. Molina, L. Munoz, A. Rela\~no, and J. Retamosa, Phys. Rev. Lett. {\bf 93}, 244101 (2004).
\bibitem{Faleiro2006} E. Faleiro, U. Kuhl, R.A. Molina, L. Mu\~noz, A. Rela\~no, and  J. Retamosa, Phys. Lett. A {\bf 358}, 251 (2006).
\bibitem{Stoeckmann1990} H.-J. St{\"o}ckmann and J. Stein, Phys. Rev. Lett. {\bf 64}, 2215 (1990).
\bibitem{Sridhar1991} S. Sridhar, Phys. Rev. Lett. {\bf 67}, 785 (1991)
\bibitem{Graef1992} H.-D. Gr\"af, H. L. Harney, H. Lengeler, C. H. Lewenkopf, C. Rangacharyulu,
A. Richter, P. Schardt, and H. A. Weidenm{\"u}ller, Phys. Rev. Lett. {\bf 69}, 1296 (1992).
\bibitem{Sirko1997} L. Sirko, P.M. Koch, and  R. Bl\"umel, Phys. Rev. Lett. {\bf 78}, 2940 (1997).
\bibitem{Blumel2001} R. Bl\"umel, P.M. Koch, and L. Sirko, Foundations of Physics {\bf 31}, 269 (2001).
\bibitem{Hlushchuk2001} Y. Hlushchuk, L. Sirko, U. Kuhl, M. Barth, and H.-J. St\"ockmann, Phys. Rev. E {\bf 63}, 046208 (2001).
\bibitem{Kottos1997} T. Kottos, U. Smilansky, Phys. Rev. Lett. {\bf 79}, 4794 (1997).
\bibitem{Kottos1999} T. Kottos, U. Smilansky, Ann. Phys. {\bf 274}, 76 (1999).
\bibitem{Pakonski2001} P. Pako\'nski, K. \.Zyczkowski, M. Ku\'s\, J. Phys. A {\bf 34}, 9303 (2001).
\bibitem{Hul2004} O. Hul, S. Bauch, P. Pako\'nski, N. Savytskyy, K. \.Zyczkowski, and L.
Sirko, Phys. Rev. E {\bf 69}, 056205 (2004).
\bibitem{Lawniczak2008} M. {\L}awniczak, O. Hul, S. Bauch, P. Seba, and L.
Sirko, Phys. Rev. E {\bf 77}, 056210 (2008).
\bibitem{Lawniczak2010}M. {\L}awniczak,  S. Bauch, O. Hul, and L.
Sirko, Phys. Rev. E {\bf 81}, 046204 (2010).
\bibitem{Hul2012} O. Hul, M.~{\L}awniczak, S. Bauch, A. Sawicki, M. Ku\'s, L.
    Sirko, Phys. Rev. Lett 109, 040402 (2012).
\bibitem{Lawniczak2014} M.~{\L}awniczak, A.~Sawicki, S.~Bauch, M.~Ku\'s, and L.~Sirko, Phys. Rev E {\bf 89,} 032911 (2014).
\bibitem{Seba1990} P. {\v S}eba, Phys. Rev. Lett. {\bf 64}, 1855 (1990).
\bibitem{Haake1991} F. Haake, G. Lenz, P. S{\v e}ba, J. Stein, H.-J. St{\"o}ckmann, and K. {\. Z}yczkowski, Phys. Rev. A {\bf 44}, R6161 (1992).
\bibitem{Shighehara1993} T. Shighehara, N. Yoshinaga, T. Cheon, and T. Mizusaki, Phys. Rev. E {\bf 47}, R3822 (1993).
\bibitem{Shighehara1994} T. Shighehara, Phys. Rev. E {\bf 50}, 4357 (1994).
\bibitem{Shighehara1996} T. Shighehara and T. Cheon, Phys. Rev. E {\bf 54}, 1321 (1996).
\bibitem{Cheon1996} T. Cheon and T. Shighehara, Phys. Rev. E {\bf 54}, 3300 (1996).
\bibitem{Weaver1995} R. Weaver and D. Sornette, Phys. Rev. E {\bf 52}, 3341 (1995).
\bibitem{Legrand1997} O. Legrand, F. Mortessagne and R. Weaver, Phys. Rev. E {\bf 55}, 7741 (1997).
\bibitem{Rahav2002} S. Rahav and S. Fishman, Nonlinearity {\bf 15}, 1541 (2002).
\bibitem{Bogomolny1998} E. Bogomolny, E. Gerland and C. Schmit, Phys. Rev. E {\bf 59}, R1315 (1998).
\bibitem{Bogomolny2001} E. Bogomolny, E. Gerland and C. Schmit, Eur. Phys. J. B {\bf 19}, 121 (2001).
\bibitem{Bogomolny2001a} E. Bogomolny, E. Gerland and C. Schmit, Phys. Rev. E {\bf 63}, 036206 (2001).
\bibitem{Bogomolny2002} E. Bogomolny, E. Gerland and C. Schmit, Phys. Rev. E {\bf 65}, 056214 (2002).
\bibitem{Exner1997} P.~Exner and P.~{\v S}eba, Phys. Lett. A {\bf 228}, 146 (1997).
\bibitem{Tudorovskiy2008} T. Tudorovskiy, R. H\"ohmann, U. Kuhl and H.-J. St\"ockmann, J. Phys. A {\bf 41}, 275101 (2008).
\bibitem{Tudorovskiy2010} T. Tudorovskiy, U. Kuhl and H.-J. St\"ockmann, New J. Phys. {\bf 12}, 123021 (2010).
\bibitem{Tudorovskiy2011} T. Tudorovskiy, U. Kuhl and H.-J. St\"ockmann, J. Phys. A {\bf 44}, 135101 (2011).
\bibitem{Rosenzweig1960} N. Rosenzweig, C. E. Porter, Phys. Rev. {\bf 120}, 1698 (1960).
\bibitem{Weyl1912} H.~Weyl, J. Reine Angew. Math. {\bf 141}, 1 (1912).
\bibitem{Bohigas1983} O. Bohigas, R.~U. Haq, and A. Pandey, in \emph{Nuclear data for science and technology}, edited by K.~H. B\"ockhoff (Reidel, Dordrecht, Netherlands, 1983), pp. 809-813.
\bibitem{Robnik1998} M. Robnik and G. Veble, J. Phys. A  {\bf 31}, 4669 (1998).
\bibitem{Lawniczak2015} M. {\L}awniczak, M. Bia{\l}ous, V. Yunko, S. Bauch, and L. Sirko, Phys. Rev. E {\bf 91}, 032925 (2015).
\bibitem{Bohigas2006} O. Bohigas and M. P. Pato, Phys. Rev. E {\bf 74}, 036212 (2006).
\bibitem{Dietz2014} B. Dietz, T. Guhr, B. Gutkin, M. Miski-Oglu, and A. Richter, Phys. Rev. E {\bf 90}, 022903 (2014).
\bibitem{Dietz2015} B. Dietz, T. Klaus, M. Miski-Oglu, and A. Richter, Phys. Rev. B {\bf 91}, 035411 (2015).
\bibitem{Dietz2015a} B. Dietz and A. Richter, CHAOS {\bf 25}, 097601 (2015).
\end{thebibliography}
\end{document}